\documentclass[aip,reprint,twocolumn]{revtex4-1}
\usepackage[english]{babel}
\usepackage{amsmath}
\usepackage{diagbox}
\usepackage{hyperref}
\usepackage{cleveref}
\usepackage{float}
\usepackage{setspace}
\usepackage{graphicx}
\Crefname{equation}{Eq.}{Eqs.}
\creflabelformat{equation}{#2#1#3}
\crefrangelabelformat{equation}{#3#1#4 to~#5#2#6}

\renewcommand\[{\begin{equation}}
\renewcommand\]{\end{equation}}

\draft 

\begin{document}


\title{Efficiency Increase in Multijunction Monochromatic Photovoltaic Devices Due to Luminescent Coupling} 



\author{Daixi Xia}
\affiliation{Department of Physics, University of Ottawa, Ottawa, Canada}

\author{Jacob J. Krich}
\affiliation{Department of Physics, University of Ottawa, Ottawa, Canada}
\affiliation{School of Electrical Engineering and Computer Science, University of Ottawa, Ottawa, Canada}


\date{\today}

\begin{abstract}
We present a multijunction detailed balance model that includes the effects of luminescent coupling, light trapping and nonradiative recombination, suitable for treatment of multijunction solar cells and photonic power converters -- photovoltaic devices designed to convert narrow-band light. The model includes both specular and Lambertian reflections using a ray-optic formalism and treats nonradiative processes using an internal radiative efficiency. Using this model, we calculate and optimize the efficiency of multijunction photonic power converters for a range of material qualities and light-trapping schemes. Multijunction devices allow increased voltage with lower current, decreasing series resistance losses. We show that efficiency increases significantly with increased number of junctions, even without series resistance, when the device has an absorbing substrate. Such an increase does not occur when the device has a back reflector. We explain this effect using a simplified model, which illustrates the origin of the decreased radiative losses in multijunction devices on substrates.
\end{abstract}


\maketitle 

\section{Introduction}
In both solar and monochromatic photovoltaics, it is well-known that a multijunction device reduces series-resistance loss compared to a single-junction equivalent, because of the low-current, high-voltage operation. High efficiency devices also rely on both photon recycling within a layer and luminescent coupling (LC) between layers, effects that are essential to accurate prediction of device performance. Monochromatic photovoltaics, also called photonic power converters (PPC's), are increasingly important components of optical power transmission systems. LC has been well studied in solar cells \cite{steiner_effects_2013,friedman_effect_2014, geisz_generalized_2015, baur_effects_2007, chan_practical_2014} and has been implemented in a drift-diffusion solver for PPC's \cite{wilkins_luminescent_2015}. There is, however, no previous detailed-balance model including LC in multijunction PPC's. Though these photovoltaic technologies are similar, there is a crucial difference between LC in solar cells and in PPC's: in solar cells, internally emitted photons can only be absorbed in the layers with lower bandgaps, so LC is one-directional; in PPC's, because all layers have the same band gap, LC is bi-directional. Previous LC models also treat only specular reflections. 

High-efficiency PPC's enable wireless power transmission isolated from electromagnetic disturbance, with applications in electric vehicles, biomedical implants, telecommunications, drones, and satellites \cite{setiawan_putra_optical_2019,wilkins_ripple-free_2019, basanskaya_electricity_2005, roeger_optically_2009}. The record-efficiency PPC is a vertical multijunction structure with 5 GaAs pn junctions coupled with tunnel diodes; it obtained an efficiency of 70\% and operating voltage of greater than 5~V at an input power density of 8 W/cm$^2$ \cite{fafard_ultrahigh_2016}. Even without consideration of series resistance, increased voltage is desirable in applications because of the removal of the need to boost the voltage. In this work, we use the term ``layers'' to refer to the active absorbing pn junctions, to avoid confusing these absorbing junctions with tunnel diodes, which are often called tunnel junctions. 

In 2001, Green used the detailed-balance formalism to show that monochromatic photovoltaic conversion can be 100\% efficient at infinite incident intensity \cite{green_limiting_2001}. Green's theory is a single-layer model, which does not capture the low current, high voltage of multi-layer operation and does not consider LC in a multi-layer device. In this work, we extend Green's theory to a multi-layer detailed-balance model and include bidirectional LC. Our model treats both specular and Lambertian top and bottom surfaces, variable incident light bandwidth, energy offset between incident light and material band gap, and nonradiative recombination by parametrizing with an internal radiative efficiency $\eta_\mathrm{int}$. We use this model to study a range of devices based on the record-efficiency PPC \cite{fafard_ultrahigh_2016}, showing the efficiency potential for future device architectures. We show that PPC efficiency increases with the number of layers, even without series-resistance loss, when the device has an absorbing substrate. This effect is not present for devices with a back reflector. We explain the origin of this effect using a simple analytic 2-layer model. This model also enables study of multijunction solar cells, with flexible application to include light trapping top and bottom surfaces.

Section \ref{sec:1layerDB} introduces the detailed balance model for a 1-layer device, including treatment of reflections from top and bottom surfaces. Section \ref{sec:multi-layer} extends this theory to the multi-layer case. Section~\ref{sec:compute-eta} shows how to transform the nonlinear equations for current $J$ as a function of voltage $V$ into a set of linear equations that allow computationally efficient extraction of $V(J)$. Section \ref{sec:application} applies the multi-layer theory to PPC's with a range of $\eta_{\mathrm{int}}$ values and light-trapping configurations, showing the intrinsic efficiency increase with number of layers when the device has an absorbing substrate. Section \ref{sec:toymodel} explains the intrinsic efficiency increase using an analytic 2-layer model.
\section{Single-layer Detailed Balance \label{sec:1layerDB}}
\begin{figure}
\begin{centering}
\includegraphics[width=\columnwidth]{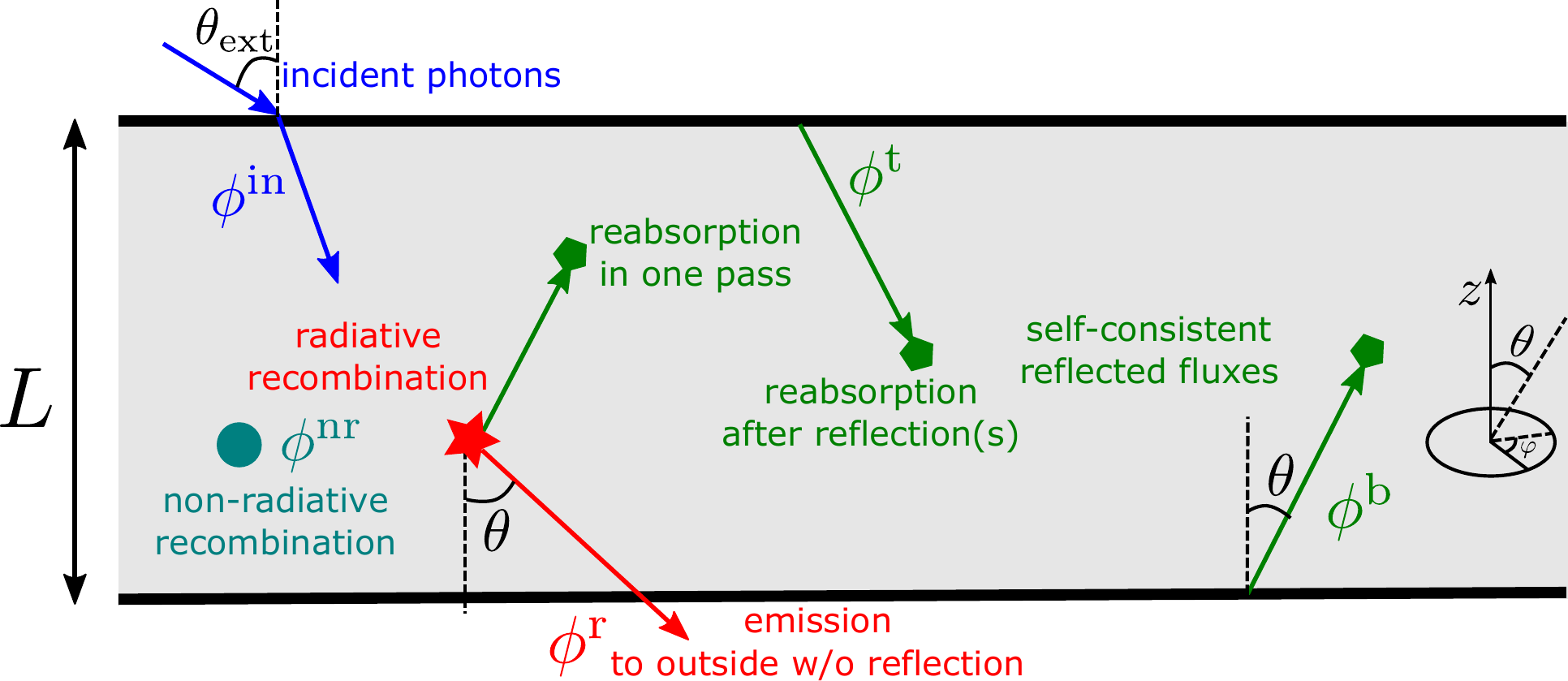}
\end{centering}
\caption{\label{fig:SJrays}Absorption and recombination events in a single-layer
cell.}
\end{figure}

We model a planar cell with infinite area and finite
thickness $L$ of the active region, as shown in Fig.~\ref{fig:SJrays}.
The detailed balance condition in this cell is
\begin{equation}
J=J^{\mathrm{in}}-J^{\mathrm{loss}}, \label{eq:1layergeneric}
\end{equation}
where $J$ is the extracted current density, $J^{\mathrm{in}}$ is the
 number of incident photons absorbed per area per time, $J^{\mathrm{loss}}$
is the net loss of current density due to electron-hole
recombination, and we set the electric charge $q=1$. $J^{\mathrm{loss}}$ is a function of the quasi-Fermi level separation $\mu$, described further below. 
We make standard detailed balance assumptions
that carrier mobilities are infinite so $\mu$ equals the applied voltage and that one photon generates one electron-hole pair \cite{shockley_detailed_1961}. The efficiency of the device is:
\[
\eta\left(\mu\right)=\frac{J\left(\mu\right)\mu}{P_{\mathrm{in}}}
\]
where $P_{\mathrm{in}}$ is the incident power density. The maximum
efficiency is obtained by optimizing $\eta$ with respect to $\mu$.  

To model both $J^{\mathrm{in}}$ and $J^{\mathrm{loss}}$, we trace the absorption
and emission of photons using ray optics. We use an angle-resolved
2D model, isotropic in the azimuth, suitable for layered structures. As shown in Fig.\ \ref{fig:SJrays},
$\theta$ is the angle between the direction of propagation of the
photons and the normal of the cell, defined from $0$ to $\pi/2$. 
When the top or bottom surfaces have specular reflections, photons with angle $\theta$ are coupled to those with angle $\pi-\theta$, 
and we label these populations with $\theta\in\left[0,\pi/2\right]$. 
Alternatively, a Lambertian surface couples photon populations of all angles to each other. 

The only challenging part of evaluating Eq.~\ref{eq:1layergeneric} lies in keeping track of the reflections off the top and bottom surfaces, both from incident and radiatively produced radiation. 
Both of these processes share the same algebraic form, and we now describe the resulting self-consistency condition including reflections.

\subsection{Self-Consistency Condition for Reflections\label{subsec:Self-Consistency-Condition-for}}

Consider the inward-directed photon fluxes from the top and bottom surfaces with angle $\theta$ and energy $E$,  
$\phi^{\mathrm{t}}\left(\theta,E\right)$,
and $\phi^{\mathrm{b}}\left(\theta,E\right)$, respectively. A self-consistency relation connects these quantities; it 
was used to model photon recycling --  reabsorption of internally emitted photons in a single-layer device -- in Ref.~\cite{balenzategui_detailed_2006} with specular reflection. Here, we exploit this self-consistency relation to model both photon recycling in single-layer structures and luminescent coupling in multi-layer structures and extend it to include Lambertian reflections.  

\begin{figure}
\begin{centering}
\includegraphics[width=\columnwidth]{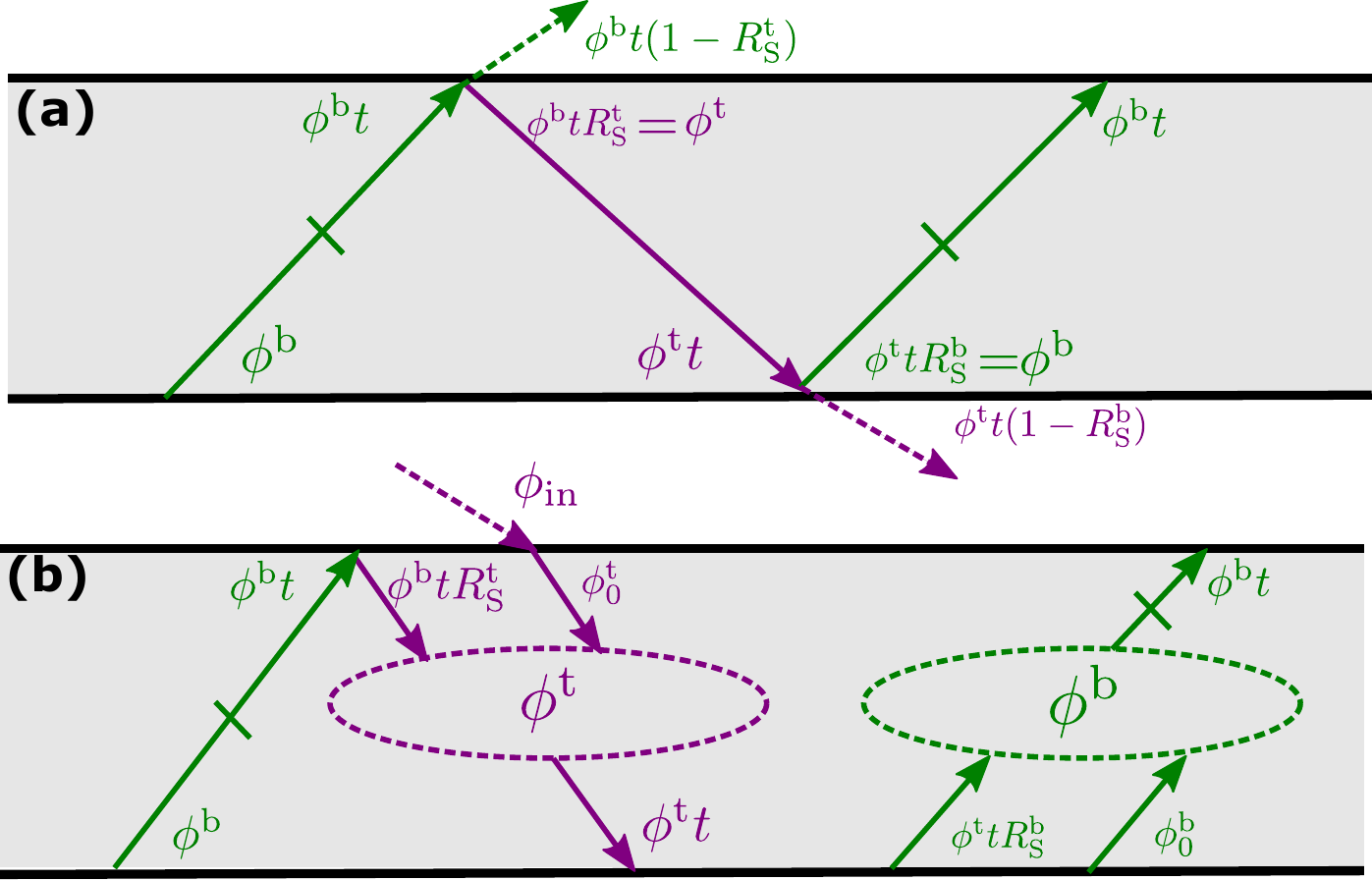}
\end{centering}
\caption{\label{fig:self-consistent}Self-consistency relation between top
and bottom inward photon fluxes, $\phi^{\mathrm{t}}\left(\theta,E\right)$,
and $\phi^{\mathrm{b}}\left(\theta,E\right)$, (a) without source terms and
(b) with source terms $\phi_{0}^{\mathrm{t}}\left(\theta,E\right)$
and $\phi_{0}^{\mathrm{b}}\left(\theta,E\right)$. 
Figure shows specular top and bottom surfaces. In the case where the source is from external radiation $\phi_\mathrm{in}$, $\phi_0^\mathrm{t}=\phi_\mathrm{in}(1-R_\mathrm{ext})$.} 
\end{figure}

Consider first specular reflections from top and bottom surfaces. As shown in Fig.~\ref{fig:self-consistent},
the inward photon flux at one surface must be equal to the flux at the
opposite surface that is transmitted through the entire cell and reflects
at the considered surface. This relation is expressed as:
\begin{align}
\phi^{\mathrm{t}}\left(\theta,E\right) & =\phi^{\mathrm{b}}\left(\theta,E\right)t\left(\theta,E\right)R_{\mathrm{S}}^{\mathrm{t}}\left(\theta,E\right)\label{eq:selfconsis_triv_top}\\
\phi^{\mathrm{b}}\left(\theta,E\right) & =\phi^{\mathrm{t}}\left(\theta,E\right)t\left(\theta,E\right)R_{\mathrm{S}}^{\mathrm{b}}\left(\theta,E\right)\label{eq:selfconsis_triv_bot}
\end{align}
where $R_{\mathrm{S}}^{\mathrm{t}}\left(\theta,E\right)$ and $R_{\mathrm{S}}^{\mathrm{b}}\left(\theta,E\right)$
are internal specular reflectivity at the top and bottom surface,
respectively. We also define $\theta$- and $E$-resolved transmittance
and absorbance:
\begin{align}
t\left(\theta,E\right) & =e^{-\frac{\alpha\left(E\right)L}{\mathrm{cos}\theta}}\\
a\left(\theta,E\right) & =1-t\left(\theta,E\right)
\end{align}
where $\alpha\left(E\right)$ is the absorption coefficient,
which we assume to be uniform in space inside the cell. Eqs.~(\ref{eq:selfconsis_triv_top}),
(\ref{eq:selfconsis_triv_bot}) can be read from Fig.\ \ref{fig:self-consistent}(a). As written, they have only the trivial solution $\phi^{\mathrm{t}}=\phi^{\mathrm{b}}=0$, 
but they express the
self-consistency condition that still applies when sources such as
external radiation or radiative recombination are included. See below,
Eqs.~\ref{eq:selfconsis_source_top}~, ~\ref{eq:selfconsis_source_bot}.
%
\begin{table*}
\begin{center}
\caption{\label{tab:phi-sols}Solution of $\phi^{\mathrm{t}}$ and $\phi^{\mathrm{b}}$ from \Cref{eq:selfconsis_source_top,eq:selfconsis_source_bot}. Depending on the source terms, these solutions appear in the text as $\phi^{\textrm{t/b}}_{\textrm{in}}$ for incident photons or as $\phi^{\textrm{t/b}}_{\textrm{lc}}$ for reflected luminescent coupling. The solution for Lambertian top and specular bottom is symmetric to the solution of specular top and Lambertian bottom, with t$\leftrightarrow$b exchanged}
\end{center}
\begin{center}
\begin{tabular}{>{\raggedleft}p{0.12\textwidth}>{\raggedright}m{0.18\textwidth}>{\centering}m{0.35\textwidth}>{\centering}m{0.3\textwidth}}
\multicolumn{2}{c}{Surface Types} & 
$\phi^{\mathrm{t}}\left(\theta,E\right)$ & $\phi^{\mathrm{b}}\left(\theta,E\right)$
\tabularnewline
\hline 
\hline 
Eq.~\ref{eq:spectop},\ref{eq:specbot} & Specular top

Specular bottom & $\frac{\phi_{0}^{\mathrm{b}}tR_{\mathrm{S}}^{\mathrm{t}}+\phi_{0}^{\mathrm{t}}}{1-t^{2}R_{\mathrm{S}}^{\mathrm{t}}R_{\mathrm{S}}^{\mathrm{b}}}$ & $\frac{\phi_{0}^{\mathrm{t}}tR_{\mathrm{S}}^{\mathrm{b}}+\phi_{0}^{\mathrm{b}}}{1-t^{2}R_{\mathrm{S}}^{\mathrm{t}}R_{\mathrm{S}}^{\mathrm{b}}}
$\tabularnewline
Eq.~\ref{eq:lambtop}, \ref{eq:lambbot} & \begin{singlespace}
Lambertian top
\end{singlespace}

Lambertian bottom & $2\mathrm{cos}\theta\frac{\Phi_{0}^{\mathrm{b}}\mathcal{T}R_{\mathrm{L}}^{\mathrm{t}}+\Phi_{0}^{\mathrm{t}}}{1-\mathcal{T}^{2}R_{\mathrm{L}}^{\mathrm{t}}R_{\mathrm{L}}^{\mathrm{b}}}$ & $2\mathrm{cos}\theta\frac{\Phi_{0}^{\mathrm{t}}\mathcal{T}R_{\mathrm{L}}^{\mathrm{b}}+\Phi_{0}^{\mathrm{b}}}{1-\mathcal{T}^{2}R_{\mathrm{L}}^{\mathrm{t}}R_{\mathrm{L}}^{\mathrm{b}}}$\tabularnewline
Eq.~\ref{eq:spectop}, \ref{eq:lambbot} & \begin{singlespace}
Specular top
\end{singlespace}

Lambertian bottom & $2\mathrm{cos}\theta\frac{\int_{0}^{\pi/2}\mathrm{sin}\theta'\mathrm{d}\theta'\left(\phi_{0}^{\mathrm{t}}tR_{\mathrm{L}}^{\mathrm{b}}+\phi_{0}^{\mathrm{b}}\right)}{1-\int_{0}^{\pi/2}\mathrm{sin}\theta'\mathrm{d}\theta'\mathrm{cos}\theta' t^{2}R_{\mathrm{S}}^{\mathrm{t}}R_{\mathrm{L}}^{\mathrm{b}}}tR_{\mathrm{S}}^{\mathrm{t}}+\phi_{0}^{\mathrm{t}}$ & $2\mathrm{cos}\theta\frac{\int_{0}^{\pi/2}\mathrm{sin}\theta'\mathrm{d}\theta'\left(\phi_{0}^{\mathrm{t}}tR_{\mathrm{L}}^{\mathrm{b}}+\phi_{0}^{\mathrm{b}}\right)}{1-\int_{0}^{\pi/2}\mathrm{sin}\theta'\mathrm{d}\theta'\mathrm{cos}\theta' t^{2}R_{\mathrm{S}}^{\mathrm{t}}R_{\mathrm{L}}^{\mathrm{b}}}$\tabularnewline
\hline 
\end{tabular}
\par\end{center}
\end{table*}
%
We also construct this self-consistency relation with one or two Lambertian
surfaces present. A Lambertian surface is an ideal diffuse scatterer that randomizes the angle of reflected
and transmitted rays \cite{green_lambertian_2002,strandberg_$jv$_2017}. Because all photons travel at random angles, the important quantity is the $\theta$-averaged
flux, $\Phi^{\mathrm{t}}\left(E\right)$, and $\Phi^{\mathrm{b}}\left(E\right)$:
\begin{align}
\Phi^{\mathrm{t}}\left(E\right) & =\int_{0}^{\pi/2}\mathrm{sin}\theta'\mathrm{d}\theta'\phi^{\mathrm{t}}\left(\theta',E\right)\label{eq:avgTop}\\
\Phi^{\mathrm{b}}\left(E\right) & =\int_{0}^{\pi/2}\mathrm{sin}\theta'\mathrm{d}\theta'\phi^{\mathrm{b}}\left(\theta',E\right)\label{eq:avgBot}
\end{align}
From the Lambert cosine law, the
$\theta$-resolved fluxes are related to the $\theta$-averaged fluxes
as \cite{strandberg_$jv$_2017}:
\begin{align}
\phi^{\mathrm{t}}\left(\theta,E\right) & =2\mathrm{cos}\theta\Phi^{\mathrm{t}}\left(E\right)\label{eq:coslawtop}\\
\phi^{\mathrm{b}}\left(\theta,E\right) & =2\mathrm{cos}\theta\Phi^{\mathrm{b}}\left(E\right)\label{eq:coslawbot}
\end{align}
Therefore, combining \Cref{eq:selfconsis_triv_top,eq:selfconsis_triv_bot,eq:avgTop,eq:avgBot,eq:coslawtop,eq:coslawbot},
\begin{align*}
\phi^{\mathrm{t}}\left(\theta,E\right) & =2\mathrm{cos}\theta\int_{0}^{\pi/2}\mathrm{sin}\theta'\mathrm{d}\theta' \phi^{\mathrm{b}}\left(\theta',E\right)t\left(\theta',E\right)R_{\mathrm{L}}^{\mathrm{t}}\left(E\right)\\
\phi^{\mathrm{b}}\left(\theta,E\right) & =2\mathrm{cos}\theta\int_{0}^{\pi/2}\mathrm{sin}\theta'\mathrm{d}\theta' \phi^{\mathrm{t}}\left(\theta',E\right)t\left(\theta',E\right)R_{\mathrm{L}}^{\mathrm{b}}\left(E\right)
\end{align*}
where $R_{\mathrm{L}}^{\mathrm{t}}\left(E\right)$ and $R_{\mathrm{L}}^{\mathrm{b}}\left(E\right)$
are $E$-resolved Lambertian reflectivities of the top and bottom surfaces, respectively.

Internal and external sources of photons add to these self-consistent fluxes, 
as shown in Fig.~\ref{fig:self-consistent}(b). We
express the inward-directed source at the top and bottom surfaces
as $\phi_{0}^{\mathrm{t}}\left(\theta,E\right)$ and $\phi_{0}^{\mathrm{b}}\left(\theta,E\right)$,
respectively. With these sources, the self-consistency
conditions at top and bottom are:
\begin{subequations}\label{eq:selfconsis_source_top}
\begin{align}
\mathrm{Specular:}\,\,\,\phi^{\mathrm{t}} & =\phi^{\mathrm{b}}tR_{\mathrm{S}}^{\mathrm{t}}+\phi_{0}^{\mathrm{t}}\label{eq:spectop}\\
\mathrm{Lambertian: }\phi^{\mathrm{t}} & =2\mathrm{cos}\theta\int_{0}^{\pi/2}\mathrm{sin}\theta'\mathrm{d}\theta' \left(\phi^{\mathrm{b}}tR_{\mathrm{L}}^{\mathrm{t}}+\phi_{0}^{\mathrm{t}}\right)\label{eq:lambtop}
\end{align}
\end{subequations}
\begin{subequations}\label{eq:selfconsis_source_bot}
\begin{align}
\mathrm{Specular:}\,\,\,\phi^{\mathrm{b}} & =\phi^{\mathrm{t}}tR_{\mathrm{S}}^{\mathrm{b}}+\phi_{0}^{\mathrm{b}}\label{eq:specbot}\\
\mathrm{Lambertian: }\phi^{\mathrm{b}} & =2\mathrm{cos}\theta\int_{0}^{\pi/2}\mathrm{sin}\theta'\mathrm{d}\theta'\left(\phi^{\mathrm{t}}tR_{\mathrm{L}}^{\mathrm{b}}+\phi_{0}^{\mathrm{b}}\right)\label{eq:lambbot}
\end{align}
\end{subequations}
where the $\theta$, $E$ dependence of all variables
has been suppressed. 
We solve for $\phi^{\mathrm{t}}\left(\theta,E\right)$
and $\phi^{\mathrm{b}}\left(\theta,E\right)$ in terms of the source
terms, for all four combinations of specular and Lambertian cases. The results
are listed in Table~\ref{tab:phi-sols}, where for simplicity of
notation, we define angle-averaged transmittance with a Lambertian
surface and angle averaged source fluxes:
\begin{align}
\mathcal{T}&=\int_{0}^{\pi/2}\mathrm{sin}\theta'\mathrm{d}\theta'2\mathrm{cos}\theta' t(\theta') \\
\Phi_{0}^{\mathrm{t/b}}  &=\int_{0}^{\pi/2}\mathrm{sin}\theta'\mathrm{d}\theta'\phi_{0}^{\mathrm{t/b}}.
\end{align}
\subsection{Current Due To Incident Photons}
With the multiple-reflection problem solved, we can express $J^\mathrm{in}$ from Eq.~\ref{eq:1layergeneric}.
We consider illumination arriving at only the top surface, hence
we take the source terms in Table \ref{tab:phi-sols} to be: 
\begin{subequations}\label{eq:incident_source}
\begin{align}
\phi_{0}^{\mathrm{t}}\left(\theta,E\right) & =\phi_{\mathrm{in}}\left(\theta,E\right)\left[1-R_{\mathrm{ext}}^{\mathrm{t}}\left(\theta,E\right)\right]\\
\phi_{0}^{\mathrm{b}}\left(\theta,E\right) & =0
\end{align}
\end{subequations}where $\phi_{\mathrm{in}}\left(\theta,E\right)$
is the number of incident photons hitting the top surface of the cell
per area per time per angle per energy, $R_{\mathrm{ext}}^{\mathrm{t}}\left(\theta,E\right)$
is the external top surface reflectivity. In experiments, the incident flux
and external reflectivity is measured in external angles, which are
related to the internal angle $\theta$ through Snell's law with a
specular surface or the Lambert cosine law with a Lambertian surface.
We express the incident flux and external reflectivity using internal
angle for simplicity of notation, and connection to experiment requires adjusting angles accordingly. We substitute Eq.~\ref{eq:incident_source}
into Table~\ref{tab:phi-sols} and obtain the current due to incident
photon absorption:
\begin{equation}
J^{\mathrm{in}}=\int_{\mathrm{hemisphere}}\mathrm{d}\Omega\int_{0}^{\infty}\mathrm{d}E\left[\phi_{\mathrm{in}}^{\mathrm{t}}+\phi_{\mathrm{in}}^{\mathrm{b}}\right]a\label{eq:Jin}
\end{equation}
where the subscript denotes that these fluxes are due only to the incident photons and their reflections, not to any internal radiative process, which will be counted in $J^\mathrm{loss}$.
\subsection{Current Due To Recombination and Photon Recycling}
We calculate $J^{\mathrm{loss}}$ using the internal per volume emission
rate and our ray-optics model to trace the reabsorption events. We calculate
the $\theta$- and $E$-resolved net loss of current and integrate
over solid angles and energy to obtain the total loss of current,
$J^{\mathrm{loss}}$. Note that for the single-layer case, this explicit ray tracing of internally emitted photons is not required, as the formulation of Green showed \cite{green_limiting_2001}. When we move to the multi-layer case, however, we must be able to track internally emitted photons and determine where they are absorbed, which requires the formalism presented here.

The $\theta$- and $E$-resolved net loss of current
is divided into three parts: (1) 
photon flux emitted out of the cell in either up or down direction,
$\phi^{r}\left(\theta,E,\mu\right)$, not including any reflections, (2) carriers lost through nonradiative
recombination, $\phi^{\mathrm{nr}}\left(E,\mu\right)$, and
(3) recycled photons absorbed after internal reflections, which enter as a negative loss,
$\phi^{\mathrm{lc}}\left(E,\mu\right)$. 
In the multi-layer case, this term will represent luminescent coupling between layers. Then, $J^{\mathrm{loss}}$ is
%
\begin{align}
J^{\mathrm{loss}}\left(\mu\right)=\int_{\mathrm{hemisphere}}\mathrm{d}\Omega\int_{0}^{\infty}\mathrm{d}E\Big [&2\phi^{\mathrm{r}}\left(\theta,E,\mu\right)+2\phi^{\mathrm{nr}}\left(E,\mu\right)\nonumber\\
&-\phi^{\mathrm{lc}}\left(\theta,E,\mu\right)\Big]\label{eq:Jloss}
\end{align}
%
Note that the factors of $2$ in Eq.~\ref{eq:Jloss} account
for both up and down propagation of photons at angle $\theta$, since we
only integrate the solid angle over a hemisphere. From here, angular integrations are always over only a hemisphere. The three terms in Eq.~\ref{eq:Jloss} all depend
on the internal radiative emission rate $S^{\mathrm{r}}$, which has dimensions of number per
time per volume per energy per solid angle \cite{wurfel_chemical_1982}:
\begin{equation}
S^{\mathrm{r}}\left(E,\mu\right)=\alpha\left(E\right)n^{2}\left(E\right)\frac{2}{h^{3}c^{2}}\frac{E^{2}}{e^{\left(E-\mu\right)/kT}-1} \label{eq:Sr}
\end{equation}
where $n\left(E\right)$ is the refractive index, $h$ is Planck's
constant, $c$ is the speed of light, $k$ is Boltzmann's constant,
$T$ is the temperature of the cell, and $\mu$ is the quasi-Fermi
level splitting in the cell. We assume that $\mu$, $T$, $n$, and $\alpha$ are spatially uniform inside
the cell, leading to spatial uniformity of $S^{\mathrm{r}}$. Eq.\ \ref{eq:Sr} implies that emission events produce photons isotropically in the cell. A photon emitted upwards
at angle $\theta$ and at position $L-z$ has a probability of $e^{-\frac{\alpha\left(E\right)z}{\mathrm{cos}\theta}}$to
escape the cell, if there are no internal reflections. On integrating
$z$ from $0$ to $L$, we obtain $\phi^{\mathrm{r}}$ in Eq.~\ref{eq:Jloss}:
\begin{align}
\phi^{\mathrm{r}}\left(\theta,E,\mu\right) & =\int_{0}^{L_{i}}e^{-\frac{\alpha\left(E\right)z}{\mathrm{cos}\theta}}S^{r}\left(E,\mu\right)\mathrm{d}z\\
 & =\frac{\mathrm{cos}\theta}{\alpha\left(E\right)}a\left(E,\theta\right)S^{r}\left(E,\mu\right).\label{eq:1layerrad}
\end{align}
%
We define the geometry factor, $g^{\mathrm{r}}$:
\begin{equation}\label{eq:phirgeo}
g^{\mathrm{r}}\left(E,\theta\right)=\frac{\mathrm{cos}\theta}{\alpha\left(E\right)}a\left(E,\theta\right),
\end{equation} 
so $\phi^{\mathrm{r}}\left(\theta,E,\mu\right) = g^{\mathrm{r}}\left(E,\theta\right) S^{r}\left(E,\mu\right)$.

We include nonradiative
recombination using $\eta_{\mathrm{int}}$, internal radiative efficiency,
which is defined as the fraction of recombination events that are radiative. 
In principle, $\eta_{\mathrm{int}}$ can depend
on position and voltage, but we assume uniform and constant $\eta_{\mathrm{int}}$ in the cell. Then, the total nonradiative loss is proportional to $L$ and $S^{\mathrm{r}}$. Similar to Eq.~\ref{eq:phirgeo}, we express $\phi^{\mathrm{nr}}$ in Eq.~\ref{eq:Jloss} as $
\phi^{\mathrm{nr}}\left(E,\mu\right)=g^{\mathrm{nr}}\left(E,\theta\right)S^{r}\left(E,\mu\right)$
where
\begin{equation}\label{eq:1layernr}
g^{\mathrm{nr}}\left(E,\theta\right)=\left(\frac{1}{\eta_{\mathrm{int}}}-1\right)L.
\end{equation}

The flux of photons that are internally emitted then reflected satisfies the self-consistency
relation as discussed in Section~\ref{subsec:Self-Consistency-Condition-for}.
We use Table~\ref{tab:phi-sols} to calculate $\phi_{\mathrm{lc}}^{\mathrm{t}}$
and $\phi_{\mathrm{lc}}^{\mathrm{b}}$, where the subscript indicates that the
source originates from the internally emitted photons. We write the
source terms as 
%
\begin{align}
\phi_{0}^{\mathrm{bdy}}\left(\theta,E\right) & =g^{\mathrm{r}}\left(E,\theta\right)S^{\mathrm{r}}\left(E,\mu\right)R^{\mathrm{bdy}}\left(\theta,E\right) \label{eq:sourceLC}
\end{align}
%
where bdy is either t or b and $R^\mathrm{bdy}$ is specular
or Lambertian reflectivity, depending on the boundary conditions chosen.
Then $\phi^{\mathrm{lc}}$ in Eq.~\ref{eq:Jloss} is
\begin{equation}
\phi^{\mathrm{lc}}=a\left(\phi_{\mathrm{lc}}^{\mathrm{t}}+\phi_{\mathrm{lc}}^{\mathrm{b}}\right).\label{eq:1layerlc}
\end{equation}
We observe from Table~\ref{tab:phi-sols} and Eq.~\ref{eq:sourceLC} that  both $\phi^{\mathrm{t}}_{\mathrm{lc}}$ and $\phi^{\mathrm{b}}_{\mathrm{lc}}$ are linear in $S^{\mathrm{r}}\left(E,\mu\right)$, allowing writing $\phi_{\mathrm{lc}}$ as $
\phi_{\mathrm{lc}} = g^{\mathrm{lc}}\left(E,\theta\right)S^{\mathrm{r}}\left(E,\mu\right)$.

Putting these results together, we express $J_{\mathrm{loss}}$ using the total geometry factor, $g\left(E,\theta
\right)$:
\begin{equation}\label{JlossGeo}
J_{\mathrm{loss}} = \int_{\mathrm{}}\mathrm{d}\Omega\int_{0}^{\infty} \mathrm{d}E  g\left( E,\theta \right) S^{\mathrm{r}}\left( E,\mu \right)
\end{equation}
where $
g\left( E,\theta \right) = 2g^{\mathrm{r}}\left( E,\theta \right) +2g^{\mathrm{nr}}\left( E,\theta \right) +g^{\mathrm{lc}}\left( E,\theta \right)$, 
and we can now evaluate all terms in Eq.~\ref{eq:Jloss} for $J^{\mathrm{loss}}\left(\mu\right)$.
With these ideas and notation established, we now extend this formalism to a multi-layer model.

\section{Multi-layer Detailed Balance \label{sec:multi-layer}}
\begin{figure}
\begin{centering}
\includegraphics[width=\columnwidth]{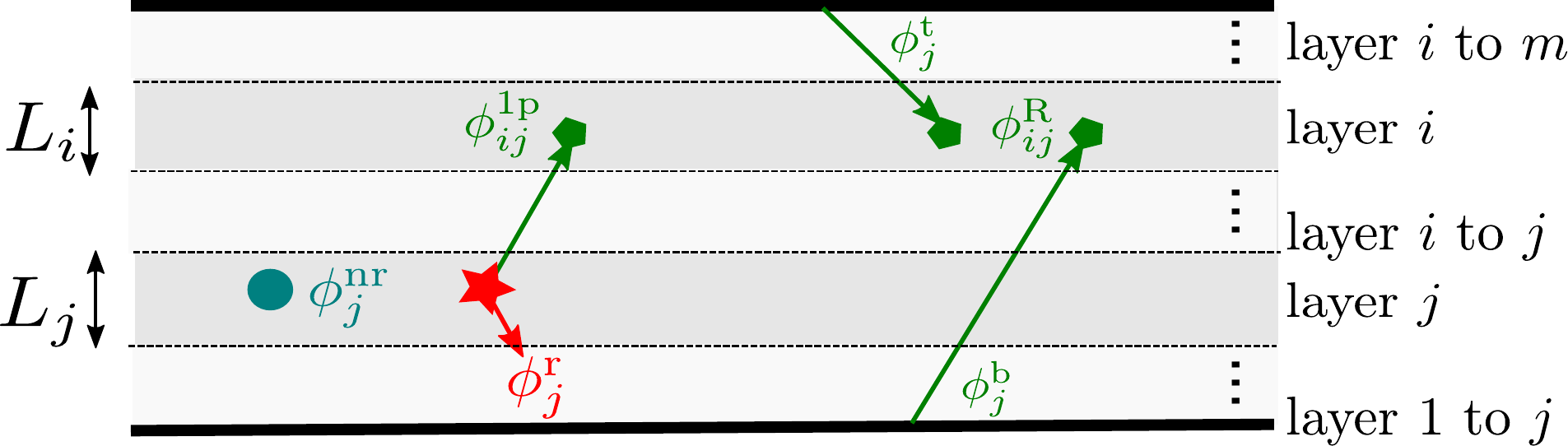}
\end{centering}
\caption{\label{fig:MJrays}Reabsorption and recombination events in a multi-layer
cell.}
\end{figure}

As shown
in Fig.~\ref{fig:MJrays}, we consider a device with $m$ vertically
stacked layers, series connected to each other. We continue to assume infinite
carrier mobility and allow each layer to have a different quasi-Fermi
level splitting, $\mu_{i}$. Each layer has
thickness $L_{i}$, absorption coefficient $\alpha_{i}\left(E\right)$
and refractive index $n_{i}\left(E\right)$, assumed to be constant within a layer. The total device thickness is $L=\sum L_i$. In most III-V devices,
refractive index does not vary significantly between layers, so in this work
we do not include reflection or refraction between layers. 

The detailed balance condition is satisfied in every layer:
\begin{equation}
J_{i}=J_{i}^{\mathrm{in}}-J_{i}^{\mathrm{loss}}\label{eq:JinMinusJloss}
\end{equation}
where $J_{i}$ is the extracted current density from layer $i$, $J_{i}^{\mathrm{in}}$
is the rate of absorption of incident photons per area in layer
$i$ and $J_{i}^{\mathrm{loss}}$ is the net loss of current density due
to recombinations in layer $i$. $J_{i}^{\mathrm{loss}}$
includes LC as a negative
loss.

We start with $J_{i}^{\mathrm{in}}$. The incident photon flux in each
layer is calculated similarly to the single layer case:
\begin{equation}
J_{i}^{\mathrm{in}}=\int_{\mathrm{}}\mathrm{d}\Omega\int_{0}^{\infty}\mathrm{d}E\left[\phi_{\mathrm{in}}^{\mathrm{t}}t_{\mathrm{t}i}+\phi_{\mathrm{in}}^{\mathrm{b}}t_{\mathrm{b}i}\right]a_{i}\label{eq:Ji_in}
\end{equation}
where we define the absorbance of each layer:
\begin{align}
a_{i}\left(E,\theta\right) & =1-e^{-\frac{\alpha_{i}\left(E\right)L_{i}}{\mathrm{cos}\theta}} \label{eq:genAbs}
\end{align}
and the transmittance through all layers between but not including $i$ and $j$ as:
\[
t_{ij}\left(\theta,E\right)=e^{-\frac{\sum_{k=i+1}^{j-1}\alpha_{k}\left(E\right)L_{k}}{\mathrm{cos}\theta}} 
\label{eq:genTrans}
\]
for $i>j+1$, and $t_{ji}=t_{ij}$. 
Note that for $\left|i-j\right|\leq1$, $t_{ij}=1$. For convenience,
we define $t_{\mathrm{t}i}$ as transmittance through all
layers above layer $i$, $t_{\mathrm{b}i}$ as transmittance through
all layers below $i$, and $t_{\mathrm{tb}}$ as the transmittance through
the whole stack. In Eq.~\ref{eq:Ji_in}, $\phi_{\mathrm{in}}^{\mathrm{t}}$
and $\phi_{\mathrm{in}}^{\mathrm{b}}$ are calculated using Table~\ref{tab:phi-sols},
with the same source terms as in Eq.~\ref{eq:incident_source}, where we replace 
$t$ with $t_\mathrm{tb}$.

We now calculate $J_{i}^{\mathrm{loss}}$ in Eq.~\ref{eq:JinMinusJloss}.
$J_{i}^{\mathrm{loss}}$ includes three terms: (1) radiative loss, $\phi_{i}^{\mathrm{r}}\left(\theta,E,\mu_{i}\right)$, (2) nonradiative loss, $\phi_{i}^{\mathrm{nr}}\left(\theta,E,\mu_{i}\right)$,
(3) luminescent coupling (LC) from layer $j$ to layer $i$, $\phi_{ij}^{\mathrm{lc}}\left(\theta,E,\mu_{j}\right)$. Similarly to the single-layer case in Eq.~\ref{JlossGeo}, we express $J_{i}^{\mathrm{loss}}$ using a geometry tensor:
\begin{equation}
J_{i}^{\mathrm{loss}} = \int_{\mathrm{}}\mathrm{d}\Omega\int_{0}^{\infty} \mathrm{d}E \sum_{j=1}^{m} g_{ij} \left(\theta,E\right) S^{\mathrm{r}}_j\left(E,\mu_j\right)\label{eq:Jloss_allterms}
\end{equation}
where $S^{\mathrm{r}}_j\left(E,\mu_j\right)$ is given by Eq.~\ref{eq:Sr} with the layer-specific $\alpha_i(E)$, $n_i(E)$, and $\mu_i$. 
%
In PPC's, all $\alpha_i$ and $n_i$ are the same in every layer, but we include the possibility of varying $\alpha$ and $n$ to to include the case of solar cells, in which each layer can have different material properties. 

We proceed to calculate the geometry tensor. Similarly to the single layer model,
\begin{equation}
g_{ij}\left( E,\theta \right) = 2\delta_{ij}g_{ij}^{\mathrm{r}}\left( E,\theta \right) +2\delta_{ij}g_{ij}^{\mathrm{nr}}\left( E,\theta \right) -g_{ij}^{\mathrm{lc}}\left( E,\theta \right) \label{eq:gij_tot}
\end{equation}
where $\delta_{ij}$ is the Kronecker delta.
Analogous to~\Cref{eq:phirgeo,eq:1layernr},
the radiative and nonradiative geometry factors in each layer are: 
\begin{align}
g_{i}^{\mathrm{r}}\left(E,\theta\right) & =\frac{\mathrm{cos}\theta}{\alpha_{i}\left(E\right)}a_{i}\left(E,\theta\right)\label{eq:phi_i_r}\\
g_{i}^{\mathrm{nr}}\left(E,\theta\right) & =\left(\frac{1}{\eta_{\mathrm{int}}^{i}}-1\right)L_{i}\label{eq:phi_i_nr}
\end{align}
We divide the LC term, $g^{\mathrm{lc}}\left(\theta,E\right)$,
into one-pass and after-reflection contributions: 
\begin{equation}
g_{ij}^{\mathrm{lc}}=g_{ij}^{\mathrm{1p}}+g_{ij}^{\mathrm{R}}.\label{eq:phi_i_lc}
\end{equation}
%
Without reflection, the geometry factor is
\begin{equation}
g_{ij}^{\mathrm{1p}}=\left(1-\delta_{ij}\right)a_{i}t_{ij}g_{j}^{r},\label{eq:phi_i_1p}
\end{equation}
where we use $(1-\delta_{ij})$ because one-pass reabsorption
within the same layer is already included in $g_{i}^{r}$. 
We obtain $\phi_{ij}^{\mathrm{R}}=g_{ij}^{R}S^r_j$ from the self-consistent reflected fluxes at the top and bottom surfaces from Table~\ref{tab:phi-sols}. For radiative events occurring in layer $j$, we take the source terms to be:
\begin{align}
\phi_{0}^{\mathrm{bdy}} & =g_{j}^{\mathrm{r}} S^{\mathrm{r}}_j t_{j\mathrm{bdy}}R^\mathrm{bdy}\label{eq:source_ij_top}, 
\end{align}
where bdy is t or b, and find $\phi_j^\mathrm{bdy}$ from Table~\ref{tab:phi-sols}. Then the absorption in layer $i$ from  radiative events in layer $j$ is:
\begin{equation}
\phi_{ij}^{\mathrm{R}}=a_{i}\left(\phi_{j}^{\mathrm{t}}t_{\mathrm{t}i}+\phi_{j}^{\mathrm{b}}t_{\mathrm{b}i}\right)\label{eq:phi_i_R}.
\end{equation}
%
%

Using~\Cref{eq:Ji_in,eq:Jloss_allterms,eq:gij_tot,eq:phi_i_r,eq:phi_i_nr,eq:phi_i_lc,eq:phi_i_1p,eq:phi_i_R,eq:source_ij_top}, 
we obtain $J_{i}$ in Eq.~\ref{eq:JinMinusJloss}. When all
layers are series connected, they share the same current density,
$J_{i}=J$, which depends on the set of quasi-Fermi levels, $\mu_{i}$.
At one $J$ value, we can solve for a set of $\mu_{i}$ values, the
sum of which gives the voltage of the device. The efficiency is then
written in terms of $J$:
\begin{equation}
\eta\left(J\right)=\frac{J\sum_{i}\mu_{i}\left(J\right)}{P_{\mathrm{in}}}\label{eq:efficiency}
\end{equation}
Optimizing with respect to $J$ gives the maximum
efficiency. 

\section{Efficient Computation of $V(J)$} \label{sec:compute-eta}

Finding $\eta(J)$ involves solving Eq.~\ref{eq:JinMinusJloss}
with $J_{i}=J$, which is a system of $m$ nonlinear equations in
$\mu_{i}$, and is computationally challenging. We can significantly reduce the computational cost by making simplifying assumptions on $\alpha\left(E\right)$ and the top and bottom reflectivities. First, we rewrite
Eq.~\ref{eq:JinMinusJloss} using Eq.~\ref{eq:Jloss_allterms}:
\begin{equation}
J_{i}=J_{i}^{\mathrm{in}}-\int_{\mathrm{}}\mathrm{d}\Omega\int_{0}^{\infty}\mathrm{d}E\left[\sum_{j=1}^{m}g_{ij}\left(\theta,E\right)S_{j}^{\mathrm{r}}\left(E,\mu_{j}\right)\right]. \label{eq:nonLinearJ}
\end{equation}
We can simplify the calculation of $J(V)$ if 
(1) $\alpha_{i}\left(E\right)$ is
zero for $E$ less than the band gap $E_{\mathrm{g}}^{i}$ and a constant for $E>E_\mathrm{g}^i$ and (2) top
and bottom reflectivity are independent of $E$. In this case, $g_{ij}(\theta,E)$ becomes $g_{ij}(\theta)$ for all $E>E_g^i$, allowing the angular integral to be separated from the energy integral. Then we
can rewrite Eq.~\ref{eq:nonLinearJ} as:
\begin{equation}
J_i=J_{i}^{\mathrm{in}}-\sum_{j=1}^{m}G_{ij}\mathcal{R}_{j}\left(\mu_j\right)\label{eq:linearJ}
\end{equation}
where
\begin{align}
G_{ij}&=\int_{\mathrm{}}\mathrm{d}\Omega g_{ij}\left(\theta\right) \\
\mathcal{R}_{j}\left(\mu_{j}\right)&=\int_{E_{\mathrm{g}}^{j}}^{\infty}\mathrm{d}ES_{j}^{\mathrm{r}}\left(E,\mu_{j}\right). \label{eq:Rj}
\end{align}
%
%
Eq.~\ref{eq:linearJ} is a linear system of equations with unknowns $\mathcal{R}_{j}$. Thus for fixed $J$ we can efficiently solve for $\mathcal{R}_j$.
We then invert $\mathcal{R}_{j}$ to find $\mu_{j}$,
using \Cref{eq:Sr,eq:Rj}, which allows us to find the external bias $V=\sum_i\mu_i$. We thus calculate $V(J)$ instead of $J(V)$. Either way, the power output is $J V$, and the maximum power point can be found numerically. 

\section{Application: Multilayer Monochromatic Conversion\label{sec:application}}

In this section, we demonstrate the use of this model by considering a set of monochromatic
devices inspired by the record-efficiency PPC device but with $m$ vertically stacked layers, varied $\eta_\mathrm{int}$, and top and bottom boundary conditions. The
effects of input power density, wavelength, linewidth, external radiative efficiency, which is different from $\eta_{\mathrm{int}}$,
absorbance and band gap were previously studied in a single-layer model  \cite{green_limiting_2001,xia_opportunities_2018} and with only one boundary condition in a multi-layer model \cite{xia_opportunities_2019}.
Here we fix those parameters as in Table~\ref{tab:fixedParams} to
approximate the record-efficiency PPC \cite{fafard_ultrahigh_2016}. We show
that devices with an absorbing substrate can improve their efficiency
by increasing the number of layers, even without series resistance.
This effect does not exist with a back reflector, which shows no improvement with increased number
of layers. In Section \ref{sec:toymodel} we present a simple model to explain these effects. 

\begin{table}
\caption{\label{tab:fixedParams}Fixed parameters for Section~\ref{sec:application}}

\centering
\begin{tabular}{lc}
\hline 
\multicolumn{1}{l}{Property} & Value\tabularnewline
\hline 
\hline 
Band gap $E_g$ \cite{adachi_optical_1999} & 1.424~eV\tabularnewline
Absorption Coefficient $\alpha$ \cite{adachi_optical_1999} & $1.151\times10^{6}$~/m\tabularnewline
Input Intensity & $8\times10^{4}$~W/m$^{2}$\tabularnewline
Input Wavelength & 830~nm (1.494~eV)\tabularnewline
Input Linewidth & 1~nm\tabularnewline
Series Resistance & 0\tabularnewline
\hline 
\end{tabular}
\end{table}

\subsection{Reflectivity Models}

For these examples, we consider a set of simple
reflectivity models to approximate different levels of light
trapping. We only consider models that are angle-dependent but not $E$-dependent
to reduce computational costs, as discussed in Section~\ref{sec:compute-eta}. The general model of Section\ \ref{sec:multi-layer} 
allows any reflectivity configuration with dependence on $\theta$ and $E$. 

\begin{table}
\caption{\label{tab:reflectivities}Top and bottom reflectivity configurations considered in this section along with icons to identify each combination.}
\centering
\begin{tabular}{l|cc}
\hline 
\multicolumn{1}{l|} {\diagbox{Bottom}{Top}} & \begin{minipage}{2 cm}\centering{Total internal \\reflection} \end{minipage}& Lambertian\tabularnewline
\hline 
\hline 
Absorbing substrate & A \includegraphics[width=0.03\textwidth]{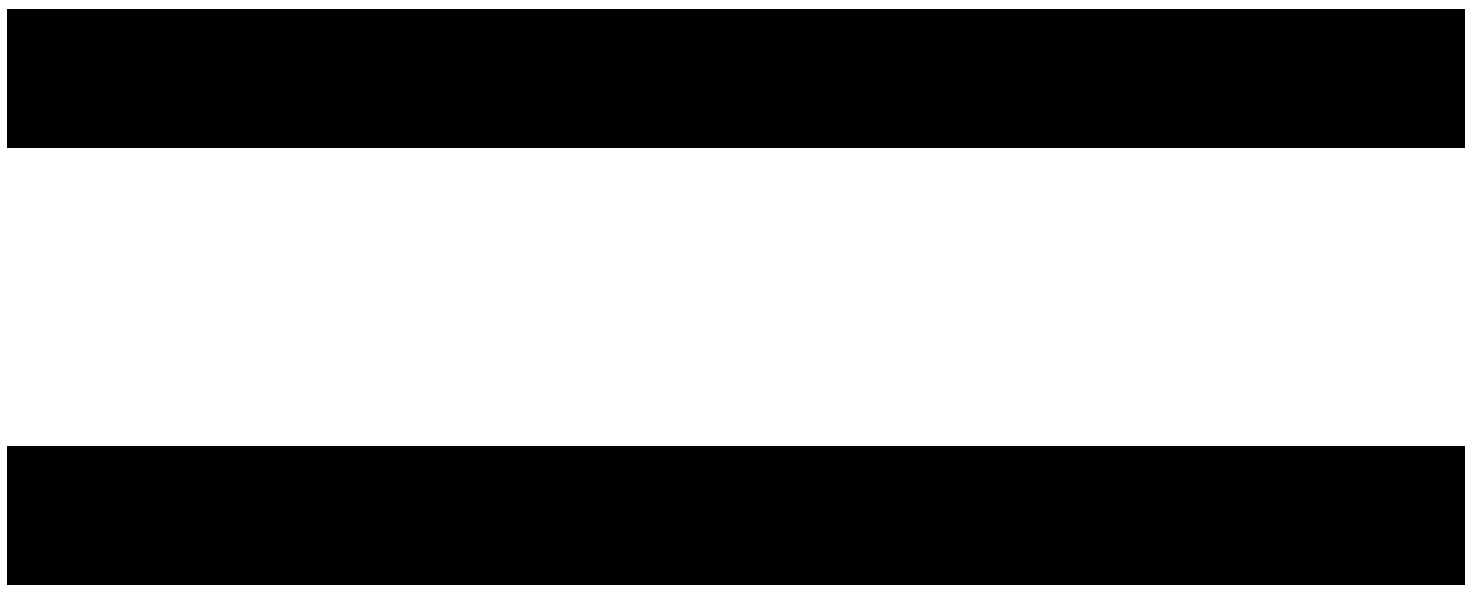} & B \includegraphics[width=0.03\textwidth]{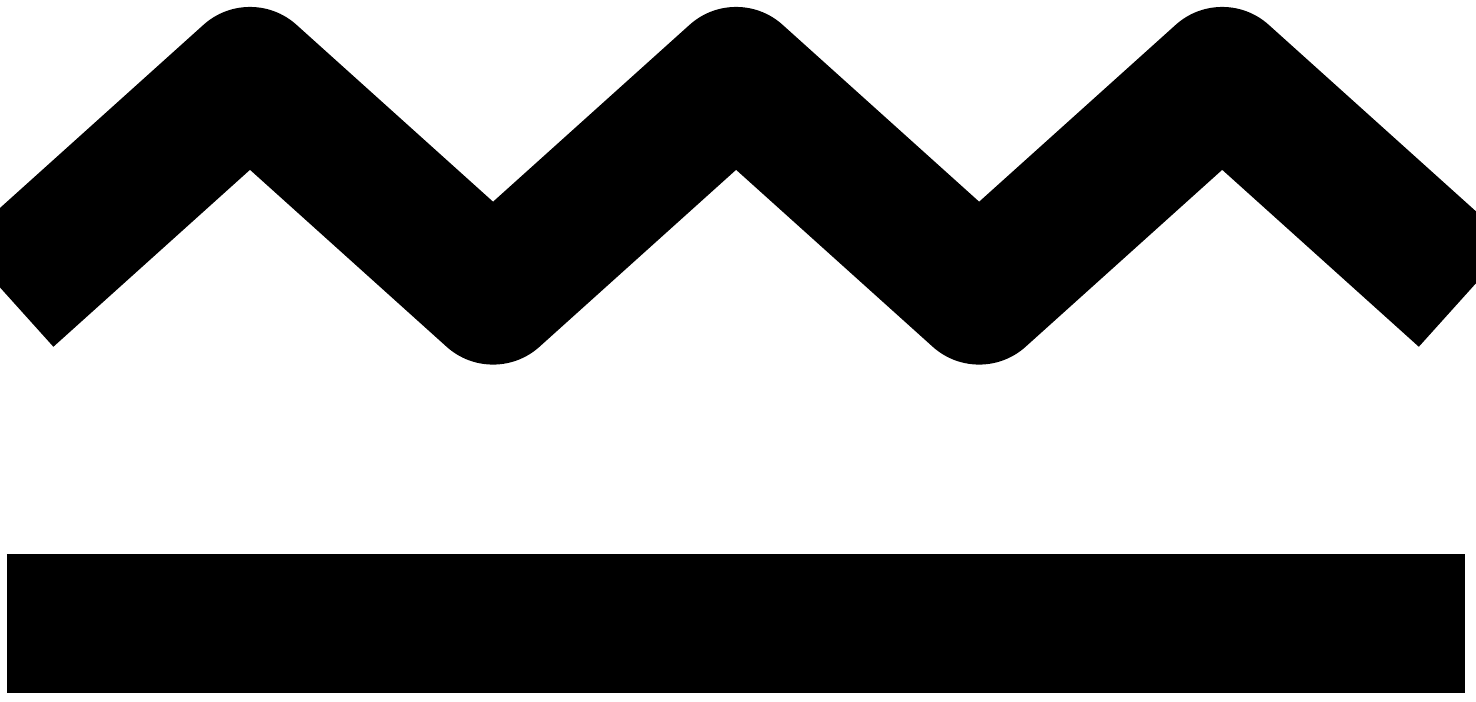}\tabularnewline
Specular mirror & C \includegraphics[width=0.03\textwidth]{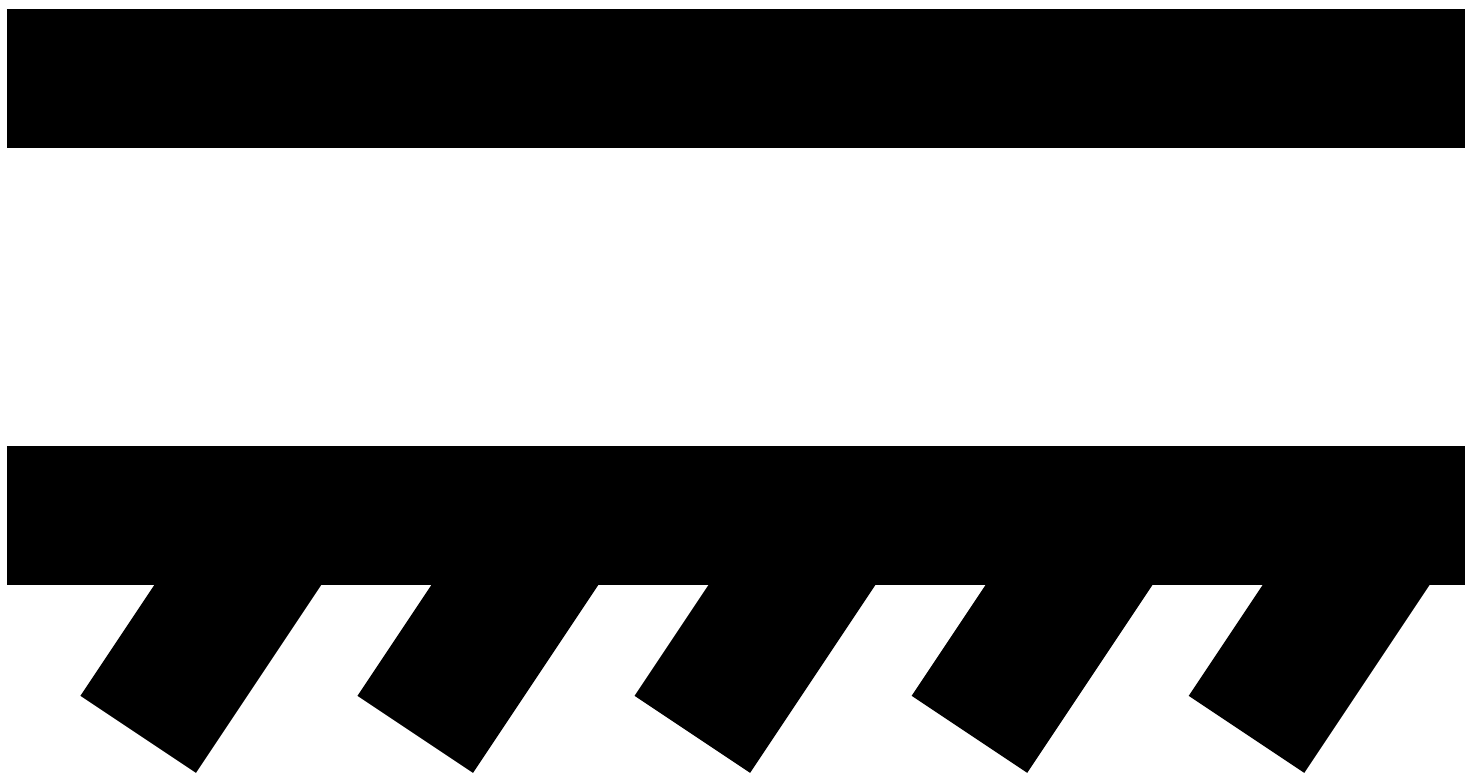} & D \includegraphics[width=0.03\textwidth]{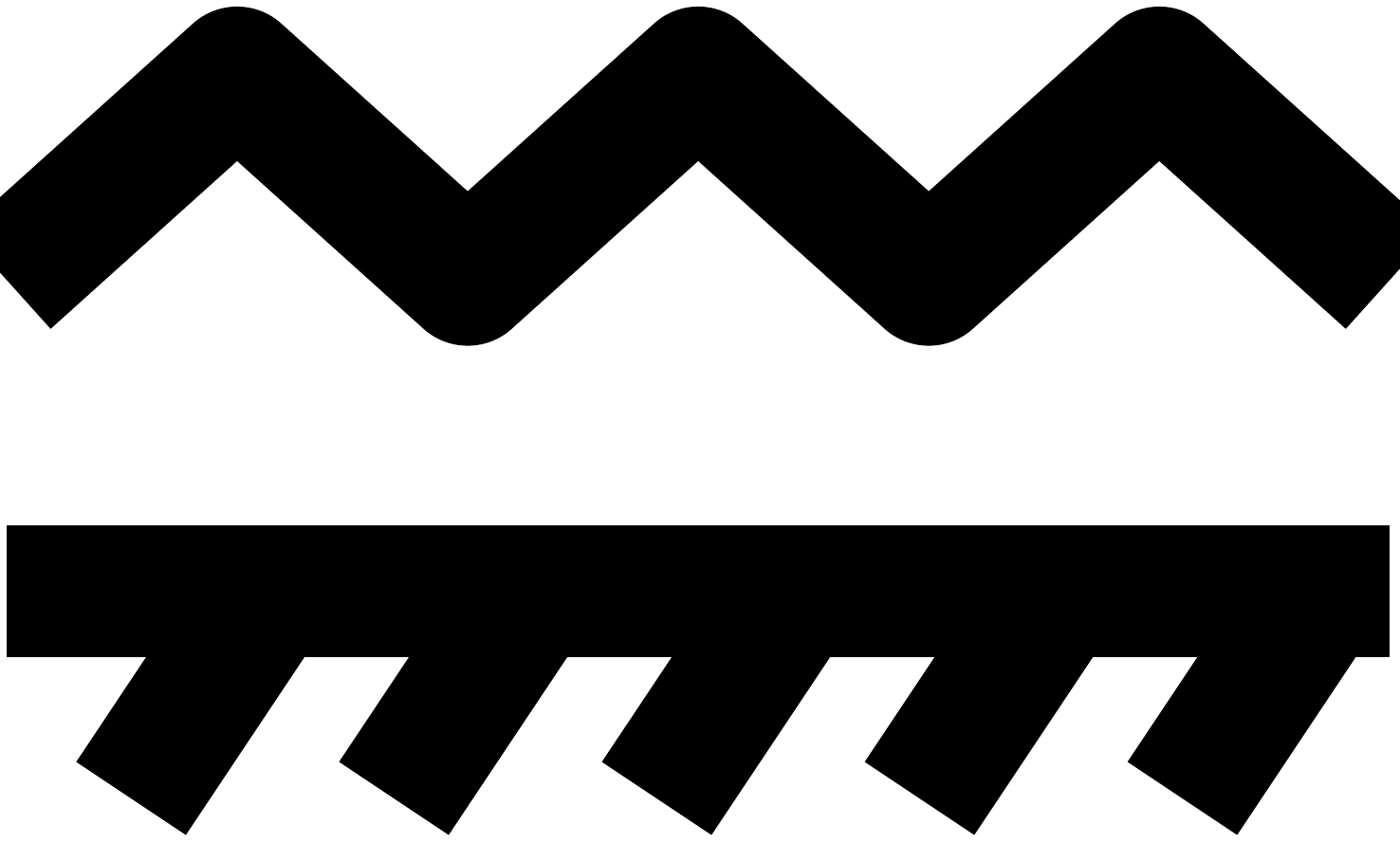}\tabularnewline
Lambertian mirror & E \includegraphics[width=0.03\textwidth]{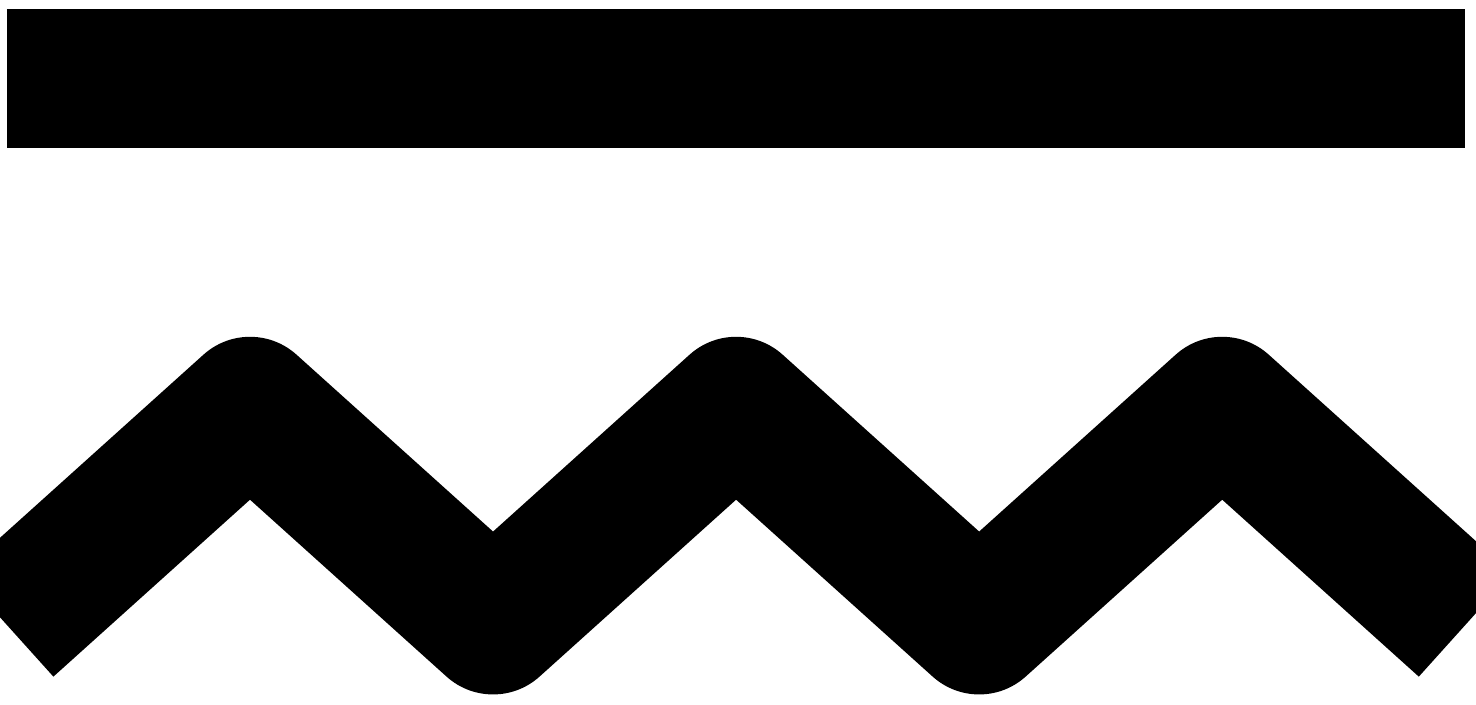}& F \includegraphics[width=0.03\textwidth]{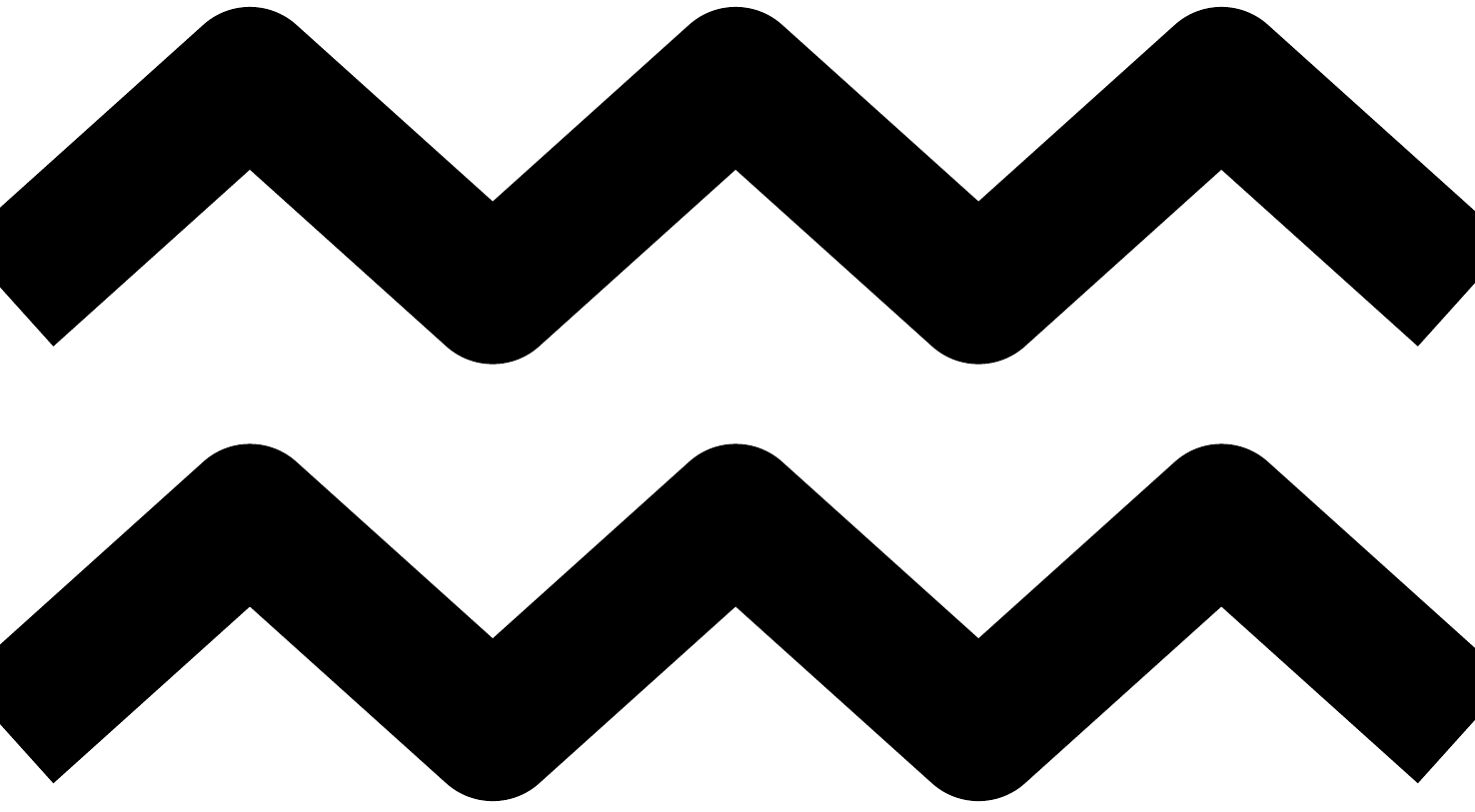}\tabularnewline
\hline 
\end{tabular}
\end{table}

We consider two models for the internal reflectivity at the top interface. In the following discussion, the top surface is an interface between air ($n_\mathrm{air}=1$) and a material with index $n$. We consider perfect transmission of external light into the sample, but there must be total internal reflection of optical modes on the inside surface. In the "Total internal reflection" model, we consider specular reflection of all incident rays with $\theta>\theta_\mathrm{c}$, 
so $R_{\mathrm{S}}^{\mathrm{t}}\left(\theta\right)$ 
is a step function that is zero for $\theta<\theta_{\mathrm{c}}$, 
and one for all larger angles. As usual, the critical angle is $\theta_{\mathrm{c}}=\sin^{-1}(n_\mathrm{air}/n)$. The second top surface we consider is an ideal Lambertian surface, which randomizes the angle of propagation for reflected and transmitted photons. An ideal Lambertian surface admits all incident photons into the cell from the exterior while reflecting
internal photons with a probability of $1-\frac{n_{\mathrm{air}}^{2}}{n^{2}}$ \cite{strandberg_$jv$_2017}. 

At the bottom surface, we consider three surface models. In the case of an
absorbing substrate, we consider all photons hitting the bottom surface
to be lost, i.e., $R_{\mathrm{S}}^{\mathrm{b}}=0$ for all angles. A specular
mirror reflects all photons back into the cell at the same angle as
incidence, i.e., $R_{\mathrm{S}}^{\mathrm{b}}=1$ for all angles. A Lambertian
mirror also reflects all photons back into the cell but at a random
angle of reflection.

\begin{figure}
\begin{centering}
\includegraphics[width=\columnwidth]{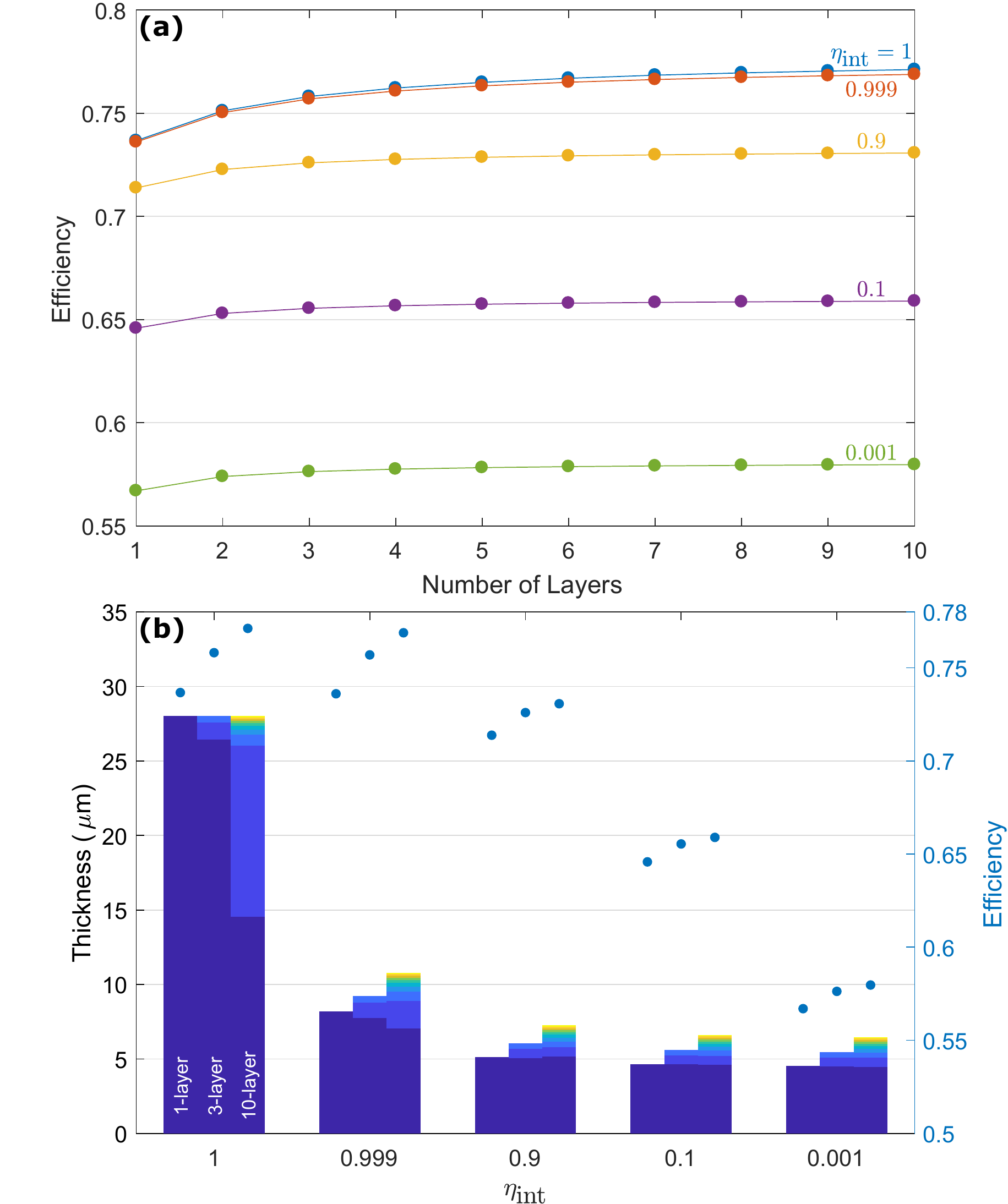}
\end{centering}
\caption{\label{fig:varyIRE}(a) Efficiency, optimized with respect to layer
thicknesses, as a function of number of layers with Configuration
A in Table~\ref{tab:fixedParams} for several values of $\eta_\mathrm{int}$. (b) Optimal layer thicknesses
of 1-, 3- and 10-layer cells (left axis, bars) and the associated efficiencies (right axis, points).}
\end{figure}

We consider all six combinations of these models for the top and bottom surface reflectivities, as listed in Table~\ref{tab:reflectivities}. 
We label the six scenarios from A to F and also include an icon for each scenario.

\subsection{Intrinsic Efficiency Increase With Number of Layers}

We first consider Configuration~A in Table~\ref{tab:reflectivities}, which has no light trapping and best represents the device of Ref.\ \cite{fafard_ultrahigh_2016}. We numerically optimize the layer thicknesses
to maximize efficiency with $\eta_{\mathrm{int}}$ values of $1$, $0.999$, $0.9$, $0.1$,
$0.001$. These values of $\eta_{\mathrm{int}}$ represent material
qualities ranging from the radiative limit to low quality.
In the radiative limit ($\eta_\mathrm{int}=1$), the optimal thickness of the full device is infinity so the cell
absorbs all incident photons. Since nonradiative
loss scales as $\left(\frac{1}{\eta_{\mathrm{int}}}-1\right)L$, there is a trade-off between nonradiative
recombination loss and transparency loss when $\eta_{\mathrm{int}}<1$, giving a finite optimal device thickness. Hence, we optimize layer
thicknesses without a constraint on total thickness for all $\eta_{\mathrm{int}}<1$
cases, while in the case of $\eta_{\mathrm{int}}=1$, we constrain the
total thickness to have a vertical one pass absorbance of $1-e^{-\alpha L}=1-10^{-14}$.

\begin{figure}
\begin{centering}
\includegraphics[width=\columnwidth]{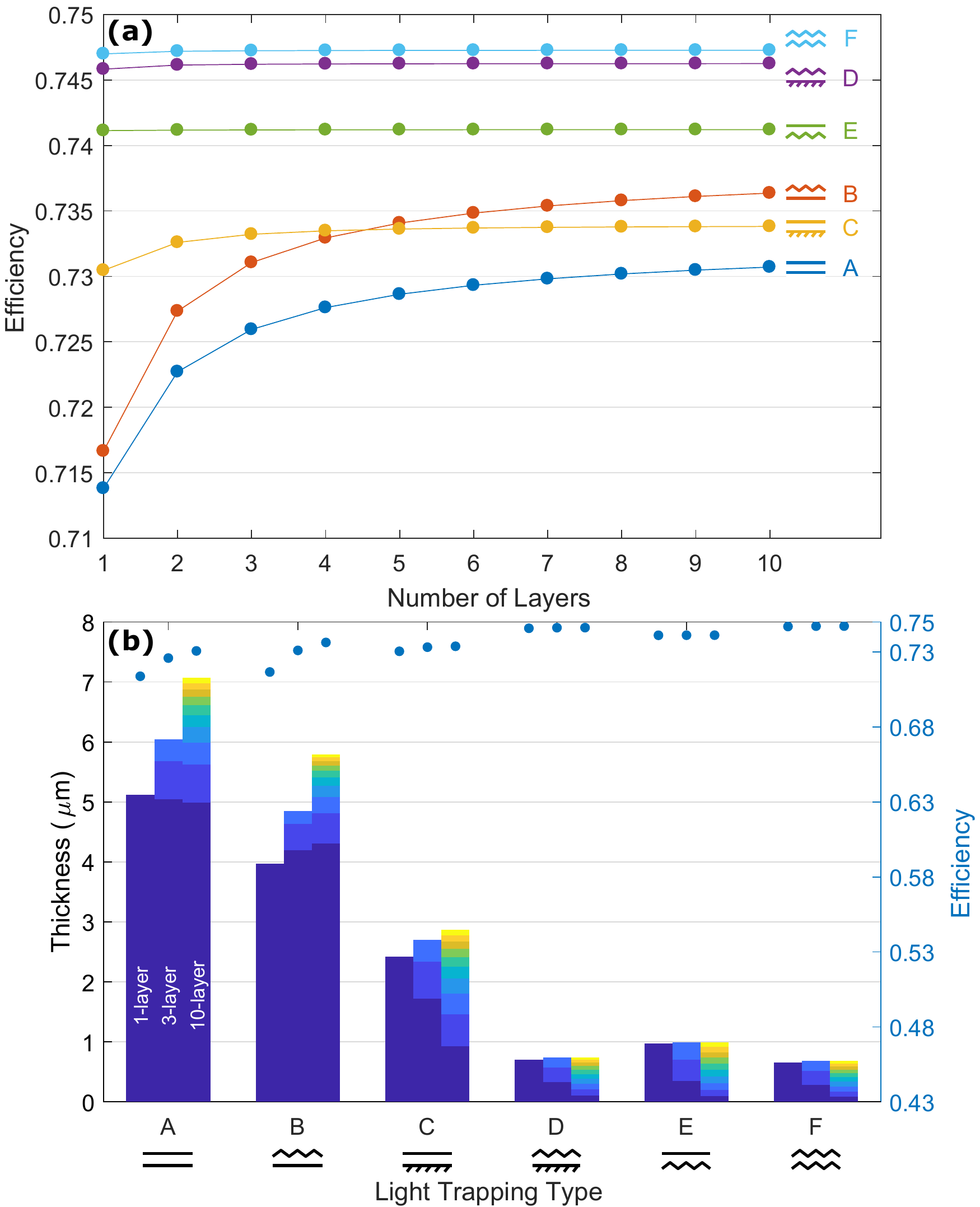}
\end{centering}
\caption{\label{fig:varySurface}(a) Efficiency, optimized with respect to
layer thicknesses, as a function of number of layers for all 6 surface
configurations in Table~\ref{tab:reflectivities}. $\eta_{\mathrm{int}}=0.9$.
Note that A, B have substrates and show the strongest intrinsic efficiency
increase with $m$. (b) Optimal layer thicknesses of 1-, 3- and 10-layer
stack.}
\end{figure}

The key results are shown in Fig.~\ref{fig:varyIRE}, giving the
optimized efficiencies and layer thicknesses for a range of $\eta_{\mathrm{int}}$ 
and $m$. As expected, efficiency decreases as $\eta_{\mathrm{int}}$ decreases.
More surprisingly, for all $\eta_{\mathrm{int}}$ values, efficiency
increases with the number of layers. In the radiative limit, efficiency
increases by 3.4\% abs. from 1 to 10 layers; at $\eta_{\mathrm{int}}=0.001$,
efficiency increases by 1.3\% abs. Efficiency is expected to improve
in multi-layer devices with series resistance, as the devices increase
voltage while decreasing current and its associated losses. In this
case, however, we see an improvement of efficiency even without series
resistance. 

We show in Fig.~\ref{fig:varySurface} efficiency at optimal layer
thicknesses for all 6 surface configurations in Table~\ref{tab:reflectivities}.
Material quality is fixed at $\eta_{\mathrm{int}}=0.9$, which represents high-quality III-V materials. Among all
configurations, case A has the lowest efficiency, because this configuration
does not include any light trapping. Efficiency is highest in Case
F with two ideal Lambertian surfaces. Only cases with absorbing
substrate (A and B) show strong efficiency increase with increasing
$m$.

A back reflector and light trapping are helpful for PPC efficiency. The record-efficiency device, however, is on a thick GaAs substrate with a full metal contact \cite{fafard_ultrahigh_2016}\cite{masson_pushing_2015}. Since the GaAs substrate is absorbing and optically thick, this architecture is equivalent to an absorbing substrate without a back reflector. In such a system, this intrinsic increase of efficiency with $m$ is an intriguing route to high efficiency, which reinforces the simpler series-resistance advantages of multi-layer devices. In order to understand the physical origin of this increased efficiency, we present a toy model of a 1- and 2-layer device, with and without a back reflector.

\section{2-Layer Model \label{sec:toymodel}}

In this section, we reduce our full model from Section~\ref{sec:multi-layer},
to a semi-analytic 2-layer model. Using this simplified model, we
demonstrate that the multi-layer device allows radiative losses into the substrate to be reduced, but that no equivalent improvement can be obtained when there is a back reflector. We consider Configurations B and F in Table~\ref{tab:reflectivities}. The simplest expectation for $m=1$ devices is
that the operating current is $J_{\mathrm{mp}}^{\left(1\right)}\simeq J^{\mathrm{in}}$,
and we call the operating voltage $V_{\mathrm{mp}}^{\left(1\right)}$.
With $m>1$, one might expect that $J_{\mathrm{mp}}^{\left(m\right)}\simeq J_{\mathrm{mp}}^{\left(1\right)}/m$
and that each layer keeps the same voltage as in the one-layer device,
giving a total voltage of $V_{\mathrm{mp}}^{\left(m\right)}\simeq m V_{\mathrm{mp}}^{\left(1\right)}$,
which would give no change in the efficiency (in the absence of series resistance). For Configuration B in Fig.~\ref{fig:varySurface}(a),
however, we see a 1.5\% (rel.) increase in efficiency on moving from
$m=1$ to $m=2$. In these devices, $mJ_{\mathrm{mp}}^{\left(m\right)}$ changes
by only 0.26\%, so the more significant increase in efficiency comes from the operating voltage. 

There is no analytic solution to $V_{\mathrm{mp}}^{\left(m\right)}$
and $J_{\mathrm{mp}}^{\left(m\right)}$. Hence, for qualitative understanding, we study the 
the short-circuit current $J_{\mathrm{sc}}^{\left(m\right)}$ and the
open-circuit voltage $V_{\mathrm{oc}}^{\left(m\right)}$, as well as the product $J_{\mathrm{sc}}^{\left(m\right)}V_{\mathrm{oc}}^{\left(m\right)}$. 
We choose configurations
with Lambertian surfaces in order to express the emission and absorption
flux using the same angle-averaged absorbance. For the same reason, we assume that there is a filter between two layers that randomizes the the angle of the transmitted light. We work in the radiative
limit and consider $n=1$, 
so there is no reflection at cell-air or cell-substrate
interfaces. We assume that absorption coefficient
is 0 for energy lower than the band gap, and constant $\alpha$ for
energy above the band gap.  For simplicity, we work in the Boltzmann approximation, in which the ``$-1'$' in the denominator of Eq.~\ref{eq:Sr} is neglected, which is valid when the internal cell voltages do not get within a few $k T$ of $E_g$. 
The detailed-balance condition, Eq.~\ref{eq:nonLinearJ}, for each layer is  then
\begin{subequations}\label{eq:simple2layer}
\begin{align}
J_{1} & =J^{\mathrm{in}}\left(1-A_{2}\right)A_{1}\left[1+R_{\mathrm{L}}^{\mathrm{b}}\left(1-A_{1}\right)\right]\\\nonumber
&-J_{0}e^{V_{1}}\left(2A_{1}-R_{\mathrm{L}}^{\mathrm{b}}A_{1}^{2}\right)
+J_{0}e^{V_{2}}A_{2}\left[A_{1}+R_{\mathrm{L}}^{\mathrm{b}}\left(1-A_{1}\right)A_1\right] \\
J_{2} & =J^{\mathrm{in}}A_{2}\left[1+R_{\mathrm{L}}^{\mathrm{b}}\left(1-A_2\right)\left(1-A_1\right)^2\right]\\\nonumber
&-J_{0}e^{V_{2}}\left[2A_{2}-R_{\mathrm{L}}^{\mathrm{b}}A_{2}^{2}\left(1-A_{1}\right)^{2}\right]
+J_{0}e^{V_{1}}\left(2A_{1}-R_{\mathrm{L}}^{\mathrm{b}}A_{1}^{2}\right)A_{2},
\end{align}
\end{subequations}
where $A$ is the angle-averaged one-pass absorbance:
\[\label{eq:angAvgAbs}
A=\frac{\int_{0}^{\pi/2}\left(1-e^{-\frac{\alpha L}{\mathrm{cos}\theta}}\right)\mathrm{cos}\theta\mathrm{sin}\theta\mathrm{d}\theta}{\int_{0}^{\pi/2}\mathrm{cos}\theta\mathrm{sin}\theta\mathrm{d}\theta}
\]
and $J_{0}$ is short-circuit radiative recombination current in one
of the up or down direction:
\[
J_{0}=\frac{2\pi}{h^{3}c^{2}}\int_{E_{\mathrm{g}}}^{\infty}\mathrm{d}EE^{2}e^{-E/kT}.
\]
The top layer absorbs the incident radiation, $A_2 J^{\mathrm{in}}$, in one-pass, and $R_{\mathrm{L}}^{\mathrm{b}}\left(1-A_2\right)\left(1-A_1\right)^2A_2J^{\mathrm{in}}$, in the second pass, where $R_{\mathrm{L}}^{\mathrm{b}}$ is the back reflectivity. The top layer radiates both out the top surface and to the bottom layer ($J_0e^{V_2}A_2$). The top layer has luminescent coupling from the bottom layer (proportional to $J_0e^{V_1}A_1$) as well as photon recycling due to the back reflector. The bottom layer receives the filtered incident radiation $(1-A_2)A_1 J^{\mathrm{in}}$ in one pass,  $R_{\mathrm{L}}^{\mathrm{b}}(1-A_2)(1-A_1)A_1 J^{\mathrm{in}}$ in the second pass, luminescent coupling in two passes from the top layer (proportional to $J_0e^{V_2}A_2$), and also has radiative emission to the top layer and the substrate. 
We can recover a 1-layer detailed balance model by setting either $A_1$ or $A_2$ to be zero. For simplicity of notation, we delete the subscript when referring to the 1-layer quantities. Using Eq.~\ref{eq:simple2layer}, we study the cases of $R_{\mathrm{L}}^{\mathrm{b}}=0$, absorbing substrate, and $R_{\mathrm{L}}^{\mathrm{b}}=1$, ideal back reflector. 

Starting with a 1-layer device, we solve for $V_{\mathrm{oc}}$ and $J_{\mathrm{sc}}$ in Eq.~\ref{eq:simple2layer} in each of the two cases:
\begin{align}\label{eq:simplevoc1}
\frac{V_{\mathrm{oc}}^{\left(1\right)}}{kT}&=\begin{cases}
\mathrm{ln}\tilde{J}+\mathrm{ln}\frac{1}{2}&\mathrm{substrate}\\ 
\mathrm{ln}\tilde{J}&\mathrm{back \enspace reflector}
\end{cases}\\
\tilde{J}^{(1)}_{\mathrm{sc}} &= \begin{cases}
A\left(\tilde{J}-2\right) &\mathrm{substrate}\\
\left(2A-A^2\right)\left(\tilde{J}-1\right) &\mathrm{back \enspace reflector}
\end{cases}
\end{align}
%
where $\tilde{J}=\frac{J^{\mathrm{in}}}{J_0}$ and $\tilde{J}_{\mathrm{sc}} =\frac{J_{\mathrm{sc}}}{J_0}$.

In 2-layer devices, the voltage of each layer at open circuit is:
\begin{align}\label{eq:simplevoc2-1}
\frac{V_{\mathrm{oc,1}}^{\left(2\right)}}{kT}&=\begin{cases}
\mathrm{ln}\tilde{J}+\mathrm{ln}\frac{\left(2-A_{2}\right)}{4-A_{1}A_{2}}
&\mathrm{substrate}\\
\mathrm{ln}\tilde{J}
&\mathrm{back \enspace reflector}
\end{cases} \\ \label{eq:simplevoc2-2}
\frac{V_{\mathrm{oc,2}}^{\left(2\right)}}{kT}&=\begin{cases}
\mathrm{ln}\tilde{J}+\mathrm{ln}\frac{\left(2+A_{1}-A_{1}A_{2}\right)}{4-A_{1}A_{2}}
&\mathrm{substrate}\\
\mathrm{ln}\tilde{J}
&\mathrm{back \enspace reflector}
\end{cases}
\end{align}
For a two-layer device at short-circuit: (1) $J_{\mathrm{sc}}=J_{1}=J_{2}$, and (2) $V_{1}+V_{2}=0$. Solving Eq.~\ref{eq:simple2layer}
with these constraints, we obtain the short-circuit current. Due to the length of the general solution, here we present only the solution evaluated at $A_1=1$ and $A_2=1/2$, which we consider further below:
\begin{equation}\label{simplejsc2}
\tilde{J}_{\mathrm{sc}}^{(2)} = \begin{cases}
\frac{\tilde{J}}{2} - \frac{7}{2\sqrt{15}} &\mathrm{substrate}\\
\frac{\tilde{J}}{2}-\frac{1}{2} &\mathrm{back\enspace reflector}
\end{cases}
\end{equation}

To understand the origin of the efficiency improvement with $m$, we compare the value of $V_{\mathrm{oc}}^{\left(m\right)}/m$ and
$J_{\mathrm{sc}}^{\left(m\right)}m$, as well as the product of
the two quantities. Since, $0\leq A_i \leq 1$, with a substrate we always
have
$
V_{\mathrm{oc,}1}^{\left(2\right)}\leq V_{\mathrm{oc}}^{\left(1\right)} \leq V_{\mathrm{oc,}2}^{\left(2\right)}
$.
The average $V_{\mathrm{oc}}^{\left(2\right)}$ with substrate can be larger or smaller
than $V_{\mathrm{oc}}^{\left(1\right)}$, depending on the values of
$A_{1}$ and $A_{2}$. 
We consider absorption-matched devices
with infinite thickness for $m=1,2$. In this case, $A=1$ for the 1-layer case, and $A_{1}=1$
and $A_{2}=1/2$ with 2 layers. Here, both the single- and 2-layer devices
absorb all incident photons. In the case of an absorbing substrate, the per-layer $V_{\mathrm{oc}}$
of the 2-layer device is strictly larger than that of the 1-layer
device. Specifically, for $\tilde{J}=10$, the average 2-layer $V_{\mathrm{oc}}$ is 6.3\% higher than the 1-layer $V_{\mathrm{oc}}$. At the same time, $m J_{\mathrm{sc}}^{\left(m\right)}$ also increases by 2.4\%. 
We thus see that the efficiency with $m=2$ is larger than with $m=1$, due to increases in both $J_\mathrm{sc}$ and $V_\mathrm{oc}$, though the $V_\mathrm{oc}$ increase is more significant. 
The product, $J_{\mathrm{sc}}^{\left(m\right)}V_{\mathrm{oc}}^{\left(m\right)}$,
increases by 8.8\%.

In the case of a substrate, the voltage difference between the top and bottom layers reduces radiative loss to the substrate. Radiative losses from the top surface are inevitable, as the device must be able to admit incident radiation. Radiative losses out the rear of the device, however, are pure losses. In a single-layer device, those losses are unavoidable at any voltage. The multi-layer device is able to reduce the impact of those losses by decreasing the internal voltage of the bottom layer, which reduces the radiation into the substrate, while allowing the upper layer to maintain a larger internal voltage. Reduced radiative loss out the bottom layer also increases the collected current $m J_\mathrm{sc}$, but this effect is smaller.

Interestingly, at higher incident flux density, both $V_\mathrm{oc}$ and $J_\mathrm{sc}$ effects are weaker. The power density used in Section~\ref{sec:application}
corresponds to $\tilde{J}\approx9\times10^{16}$. With this value, in the toy model $V_{\mathrm{oc}}^{\left(m\right)}/m$
increases by $0.26\%$, while $mJ_{\mathrm{sc}}^{\left(m\right)}$
stays the same within machine precision. In the full model of Section\ \ref{sec:multi-layer} with $n=1$ and $\eta_\mathrm{int}=1$, the improvement of maximum power on moving from $m=1$ to $m=2$ for Configuration B is $0.22\%$, in good agreement with the simpler model. With GaAs refractive index, $n=3.64$, the improvement is 2.0\%, similar to that shown in Fig.~\ref{fig:varySurface}(a), showing that the increased top-surface light trapping increases the importance of this effect.

In contrast, for the case with a back reflector, the voltage of each layer in a 2-layer device is analytically equal to the voltage of the single-layer device in all $A_1$ and $A_2$ values. And $2J_\mathrm{sc}^{(2)}$, evaluated at the absorption matched case of $A_1=1$ and $A_2=1/2$, is also analytically equal to $J_\mathrm{sc}^{(1)}$ at $A=1$. This matches our observation in the full model that efficiency does not increase with number of layers in Configuration~F.

The highly nonradiative case,  $\eta_{\mathrm{int}}\ll 1$, is also amenable to analytic treatment and shows a similar increase of efficiency with $m$.

We have generalized monochromatic detailed balance models \cite{green_limiting_2001,xia_opportunities_2018} to multi-layer devices, including: (1) bi-directional luminescent coupling, (2) nonradiative recombination parametrized by $\eta_{\mathrm{int}}$, (3) mixed specular and Lambertian surface reflection. We observe an intrinsic increase of efficiency with number of layers, independent of series resistance, for PPC's on absorbing substrates. This intrinsic increase is not present when the device has a back reflector.
Other than the well-known benefits of multi-layer PPC's, such as high voltage and low series resistance, we have discovered another mechanism for efficiency increase with number of layers, which further encourages future multi-layer designs of PPCs.

\section*{Acknowledgement}
We acknowledge helpful conversations with Matthew M. Wilkins. This work was supported by NSERC grant number STPGP 494090 and Ontario Early Researcher Award ER17-13-019. The data that support the findings of this study are available from the corresponding author upon reasonable request.

\bibliography{mjdb_method_paper_aip}

\end{document}